\documentclass{aa51}
\usepackage{psfig}
\usepackage{times}
\newcommand{\degree}{\mbox{$^{\circ}$}}               
\newcommand{\micron}{\mbox{\,${\mu}$m}}               
\newcommand{\Msolar}{\mbox{\,$M_{\odot}$\/}}          
\newcommand{\Lsolar}{\mbox{\,$L_{\odot}$\/}}          
\newcommand{\Mjup}{\mbox{\,$M_{\rm Jup}$\/}}          
\newcommand{\HII}{\mbox{H\,{\footnotesize II}}}       
\newcommand{\magnit}[2]{\mbox{$\mbox{\rm #1}^{\mbox{\rm\tiny m}}%
     \!\!\!.\!\,\, \mbox{\rm #2}$}}                   
\newcommand{\magap}[1]{\mbox{$\mbox{\rm #1}^{\mbox{\rm\tiny m}}$}} 
\newcommand{\filter}[1]{\mbox{\it #1\/}}              
\newcommand{\colour}[2]{\mbox{(\it #1$ - $#2\/})}     
\newcommand{\oversim}[2]{\lower0.5ex\vbox{\baselineskip=0pt\lineskip=0.2ex
     \ialign{$\mathsurround=0pt #1\hfil##\hfil$\crcr#2\crcr\sim\crcr}}} 
\newcommand{\etal}{\mbox{\hbox{et\,al.}}}         
\newcommand{\idest}{\mbox{\hbox{\it i.e.,}}}          
\begin{document}
\title{The Eagle's EGGs: fertile or sterile?}
\author{Mark J. McCaughrean \and Morten Andersen}
\offprints{Mark McCaughrean}
\institute{Astrophysikalisches Institut Potsdam,
An der Sternwarte 16, 14482 Potsdam, Germany \\
\email{mjm@aip.de, mortena@aip.de}
}
\date{Received, accepted}
\abstract{We present a deep, high spatial resolution (0.35 arcsec FWHM), 
near-infrared (1--2.5\micron) imaging survey of the Eagle Nebula, M\,16, made 
with the VLT, centred on the famous elephant trunks. We compare these data 
with the existing HST optical images to search for evidence of ongoing or 
recent star formation in the trunks, and in particular in the 73 small 
evaporating gaseous globules (EGGs) on their surface. We find that two of 
the three HST trunks have relatively massive YSOs in their tips. Most 
of the EGGs appear to be empty, but some 15\% of them do show evidence for 
associated young low-mass stars or brown dwarfs: in particular, there is a 
small cluster of such sources seen at the head of the largest trunk.
}
\maketitle
\section{Introduction}
The life history of a star appears to be determined both by `nature', imprints 
received prior to birth, and `nurture', influences from the environment 
surrounding it in the early years after birth. Crucial amongst these external 
influences is the presence of massive stars: their intense ionising flux and 
strong winds can turn the common parental molecular cloud into an \HII{} region 
and thus terminate star formation, as well as evaporate circumstellar disks 
around low-mass stars and prevent planet formation (Hollenbach, Yorke, \&
Johnstone 2000). On the other hand, massive stars can also lead to radiative 
implosion of surrounding molecular cores, initiating secondary star formation 
(Larosa 1983; Bertoldi 1989; Lefloch \& Lazareff 1994).

A famous example of the effects wrought by massive stars is seen in the Eagle 
Nebula, M\,16. OB stars in the young stellar cluster NGC\,6611 (Hillenbrand 
\etal{} 1993) are photoionising surrounding molecular material, leading in one 
region to the creation of three elongated columns or `elephant trunks', as 
detailed in the iconic Hubble Space Telescope images of Hester \etal{} (1996; 
hereafter H96). H96 identified 73 small (typically $\sim$\,0.5 arcsec or 
1000\,AU diameter) protrusions on or near the surface of the trunks, which 
they dubbed EGGs, for evaporating gaseous globules. Noting a few apparent 
associations with stars seen in the HST and/or near-infrared images 
(Hillenbrand \etal{} 1993), H96 proposed that the EGGs are created as the 
NGC\,6611 ionisation front sweeps over dense condensations of molecular 
material harbouring young stars, \idest{} they might also be considered as 
`eggs'. These embedded stars would be exposed prematurely as the surrounding 
EGG is evaporated, thus removing the reservoir for future accretion and 
perhaps also the circumstellar disk. H96 suggested that if most stars form 
under the influence of massive stars, this influence might then ultimately 
determine important results such as the stellar initial mass function and the 
fraction of stars with planetary systems. 

However, missing in this hypothesis is direct observational evidence that the 
M\,16 columns and EGGs indeed contain a substantial population of young stars. 
As the optical images of H96 were not able to probe the dusty interiors of the 
columns, longer wavelength data are required. The region has been surveyed in 
the near-infrared by Walsh \& White (1982), Chini \etal{} (1992), Hillenbrand 
\etal{} (1993), Currie \etal{} (1996), and McCaughrean (1997), in the 
mid-infrared by Pilbratt \etal{} (1998), and in the millimetre by Pound (1998) 
and White \etal{} (1999). However, none of these studies had the required 
combination of high spatial resolution, sensitivity, and field-of-view to 
provide a detailed view of possible ongoing star formation in the trunks, or 
to carry out a statistically meaningful survey for embedded sources in the 
EGGs.

Two recent near-infrared imaging studies (Sugitani \etal{} 2002, hereafter S02; 
Thompson, Smith, \& Hester 2002, hereafter TSH02) have revisited the region 
with better spatial resolution and sensitivity, and both describe evidence for 
massive star formation in the tips of two of the trunks, as also seen in the 
previous images of Hillenbrand \etal{} (1993) and McCaughrean (1997). However, 
neither study directly addressed the question of star formation in the EGGs.
We have used the VLT to make a new near-infrared survey of a large region 
containing NGC\,6611, the elephant trunks, and their EGGs, with sufficient 
spatial resolution ($\sim$\,0.35 arcsec FWHM) and sensitivity 
(\filter{K$_s$}$>$\magap{20}) to probe EGGs individually. Here we present just 
the data covering the elephant trunks and our essential findings concerning the 
EGGs: the entire data set can be seen in McCaughrean \& Andersen (2001), and 
a detailed analysis will be given in a subsequent paper.

\section{Observations}
The data were obtained with the facility near-infrared camera/spectrograph, 
ISAAC, on the 8.2-m VLT UT1 Antu in service mode (ESO program 67.C-0595) 
on April 10 and May 8--10 2001. The ISAAC short-wavelength camera has a 
1024$\times$1024 pixel HgCdTe HAWAII array covering 2.5$\times$2.5 arcmin 
at 0.147 arcsec/pixel. A slightly-overlapping 4$\times$4 position mosaic was 
made to cover a $\sim$\,9$\times$9 arcmin region centred on the elephant trunks 
and including much of NGC\,6611: the inner 2$\times$2 mosaic was imaged 
repeatedly to go deep on the most critical region, while the outer part of 
the mosaic was covered less often, to provide sky data and a broader survey. 
Data were taken in three broad-band filters \filter{J$_s$} (1.24\micron), 
\filter{H} (1.65\micron), and \filter{K$_s$} (2.16\micron). A standard 
observing block involved imaging each position in the central 2$\times$2 
mosaic three times and for 100 seconds each time (10 coadds of 10 seconds); 
the outer ring of 12 positions in the 4$\times$4 position mosaic was imaged 
once for 50 seconds (5 coadds of 10 seconds). 

Data reduction was standard. For any given image, a blank sky frame was 
constructed by making a median stack of the other 23 images within the same
block. A clean sky frame was always achieved, despite the extremely crowded 
nature of the region. This sky frame was subtracted from the image and 
the result then flat-fielded using differential twilight flats (\idest{} 
bright sky minus faint sky). Corrections were made for the small geometric 
distortion in ISAAC before the images were aligned, intensity normalised, and 
co-added to form three final mosaics, one per filter.
 
In \filter{H} and \filter{K$_s$}, data from only the better of two observing 
blocks were used, while in \filter{J$_s$}, data from four equal-quality blocks 
were used. Thus for the deepest, central section of the mosaics discussed here, 
the effective integration time per pixel is 300 seconds in \filter{H} and 
\filter{K$_s$}, and 1200 seconds in \filter{J$_s$}, the longer integration
time at \filter{J$_s$} in part compensating for higher extinction in the EGGs 
at the shorter wavelength. Spatial resolution in the coadded mosaics is 0.38, 
0.36, and 0.33 arcsec FWHM in \filter{J$_s$}, \filter{H}, and \filter{K$_s$}, 
respectively. Photometric calibration via faint standard stars yields detection 
limits (defined as a star whose brightest pixel is detected at 3$\sigma$ above 
the local noise) at \filter{J$_s$}, \filter{H}, and \filter{K$_s$} of 
\magnit{22}{6}, \magnit{21}{3}, and \magnit{20}{4}, respectively. In regions 
of bright nebular emission, these limits are somewhat reduced. 

\begin{figure*}
\centerline{\psfig{figure=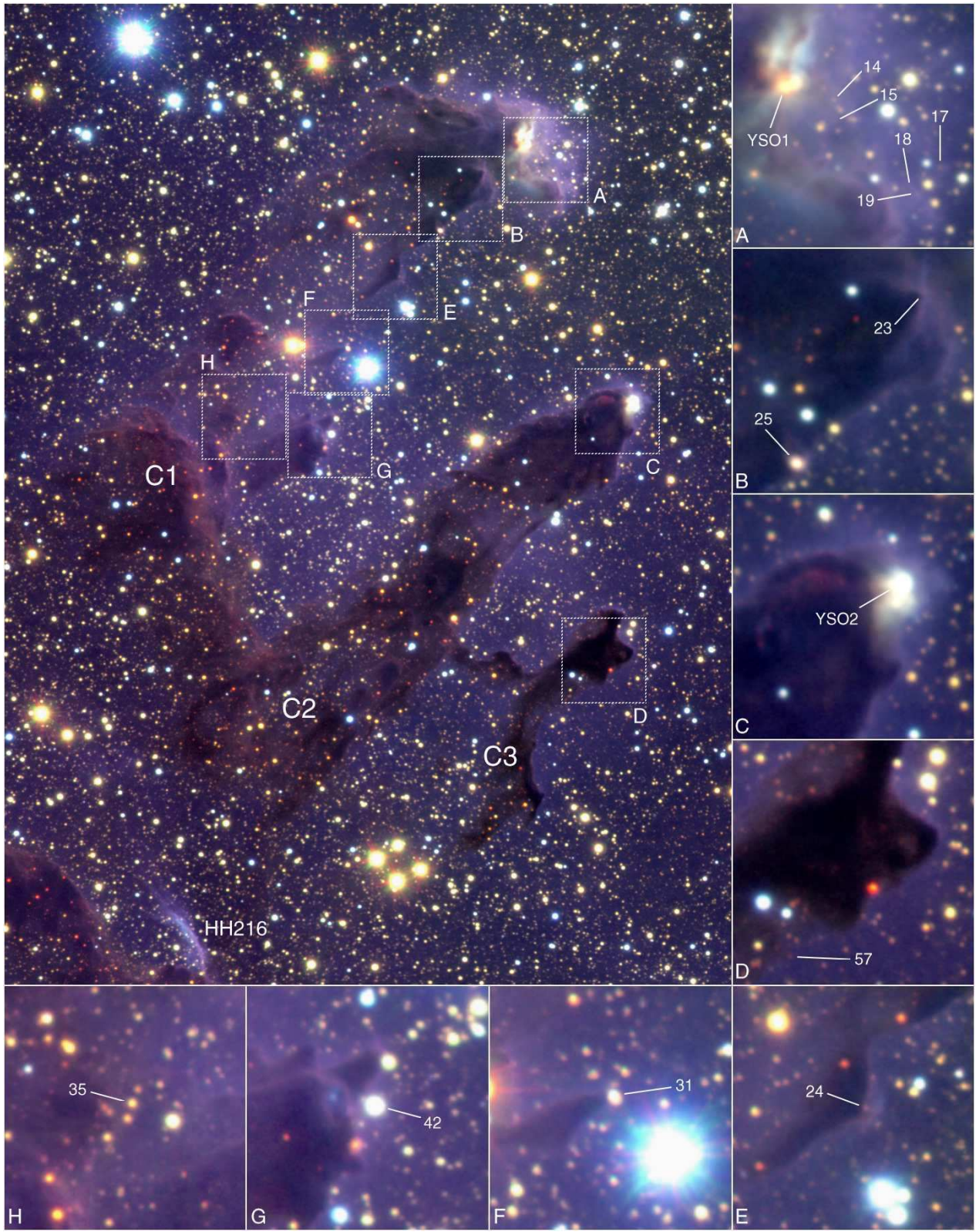,height=21.2cm}}
\caption{True-colour near-infrared (1--2.5\micron) image of the M\,16
elephant trunks. The \filter{J$_s$} data are shown as blue, \filter{H} as 
green, and \filter{K$_s$} as red. The cube root of the intensities were taken 
to compress the dynamic range before normalising and combining the three 
mosaics. The main image is centred at 
18$^{\rm h}$\,18$^{\rm m}$\,52.7$^{\rm s}$, 
$-13$\degree\,50$^{\prime}$\,09$^{\prime\prime}$ (J2000.0)
and covers 158$\times$214 arcsec (1.5$\times$2.0\,pc at 1.9\,kpc);
north is up, east left. The subimages have been magnified by a factor of 
2.9 and each covers 18.5$\times$18.5 arcsec (0.17$\times$0.17\,pc).
Labels mark EGGs from H96 found to be associated with point sources
as described in the text; E23, an EGG with no near-infrared point source, but 
thought to contain an embedded protostar driving a collimated jet; YSO1 and 
YSO2, massive sources in the tips of C1 and C2, respectively; and HH\,216, an 
optically-visible Herbig-Haro object. Due to the large dynamic range, some of 
the very faintest sources are not easily seen in these subimages, but can be 
seen in the original data.
}
\label{fig:trunks}
\end{figure*}

Figure~\ref{fig:trunks} shows the region centred on the trunks in a
three-colour composite. This region and the complete survey mosaic can be 
seen at full resolution in McCaughrean \& Andersen (2001). 

\section{Results}
The contrast between the HST optical emission-line image of H96 and the VLT 
near-infrared broad-band image is striking. The HST image is dominated by the 
three gas and dust columns: the VLT image shows not only these, but also a 
huge number of stars towards, in, and beyond M\,16. Stars seen in the 
VLT image as blue are in M\,16 or in the foreground, but the great majority 
of the stars are fainter and yellow, and are background field stars seen along
the line-of-sight through the bulge towards the inner galaxy 
($l,b$\,$\sim$\,17\degree,+1\degree), through the molecular material out 
of which M\,16 was created. An infrared colour-colour diagram derived from
our full data set and containing some 5$\times$10$^4$ stars shows this 
background screen to have an extinction of A$_V$\,$\sim$\,\magap{10}. There 
are also a number of orange and red sources associated with the columns and 
thus seen through additional dust extinction. However, simple visual 
examination cannot distinguish between stars embedded in the trunks or 
background stars seen through them.

The three columns are also considerably changed in the near-infrared, as noted 
and discussed by S02 and TSH02. The heads of the three columns are slightly 
reduced in size with respect to the optical images, but all retain significant 
opaque cores. The largest column (C1) disappears almost completely along much 
of its length below the head (cf.\ Pound 1998), and the middle trunk (C2) is 
also broken in the middle: only the smallest trunk (C3) remains continuous. 

\subsection{Star formation in the columns}
The immediate goal of our VLT experiment was to enable a sensitive search for 
embedded young stars in the EGGs.
To characterise any such objects seen in our images, we 
use the pre-main sequence models of Baraffe \etal{} (1998) at $\leq$1\Msolar{} 
and Palla \& Stahler (1999) at higher masses. To generate magnitudes for the 
ISAAC \filter{J$_s$} and \filter{K$_s$} filters, we have 
interpolated those available for the standard \filter{J}, \filter{H}, and 
\filter{K} filters according to band centre, a justifiable approximation
given the close similarity of the two filter sets. We assume a distance of 
1.9\,kpc to M\,16 and R=3.75 for its extinction law (Hillenbrand \etal{} 
1993), and a generic age of 1\,Myr for all putative young embedded sources.
Under these assumptions, our VLT detection limits correspond to a limiting 
mass of $<$0.01\Msolar{} (10\Mjup{}) in all three filters, assuming
no extinction. However, since we are searching for sources embedded in the 
EGGs, we calculate that a 0.075\Msolar{} source at the star:brown dwarf 
boundary can be detected through up to A$_V$\,$\sim$\,\magap{16}, \magap{23}, 
\and \magap{30} of extinction at \filter{J$_s$}, \filter{H}, and 
\filter{K$_s$}, respectively, for an R=3.75 extinction law (Cardelli, Clayton, 
\& Mathis 1989). Assuming an embedded star lies in the centre of its EGG, 
then the 0.5 arcsec median diameter of an EGG converts to a column of 
7$\times$10$^{15}$\,cm to the source. To yield an extinction of 
A$_V$\,=\,\magap{30}, the mean volume density would then have to be 
$\sim$10$^7$\,cm$^{-3}$, $\sim$\,50 times greater than is measured for the 
heads of the trunks (White \etal{} 1999; Pound 1998). Therefore, despite
a likely above average density in the EGGs, the present data appear to have 
the sensitivity required to probe most of the EGGs to the brown dwarf 
limit and below.

Without classification spectroscopy, only crude mass estimates can be made for 
detected sources, by placing them in the infrared colour-magnitude diagram and 
dereddening until they intercept the model 1\,Myr isochrone. As some sources 
will show excess long-wavelength emission in the colour-colour diagram 
indicative of a circumstellar disk, the \filter{J$_s$} versus \colour{J$_s$}{H} 
colour-magnitude diagram is generally used for the mass estimation. It is 
clear that this method is approximate at best, and the results presented in 
Table~1 should be taken only as an indication of the likely masses of the 
embedded sources, in lieu of further observations.

\subsubsection{Massive YSOs in the columns}
The tips of both C1 and C2 show evidence for bright embedded young stellar 
objects (YSOs). Both were detected in earlier infrared images (Hillenbrand 
\etal{} 1993; McCaughrean 1997), and have been described in detail by S02 and 
TSH02. The source in C1 (here YSO1; P1 of S02; M16ES-1 of TSH02) is very red 
and sits at the southern tip of a complex extended nebula bisected by a lane 
of extinction associated with one of the more prominent `towers' on the head 
of C1. It is also detected at 7--11\micron{} by MSX (Price \etal{} 2001), 
12--18\micron{} by ISO (Pilbratt \etal{} 1998), and as a bright hard X-ray 
source by Chandra (Gagn\'e, personal communication).

Assuming the \filter{J$_s$} and \filter{H} fluxes (see Table~1) of YSO1 to be 
photospheric, we deduce an extinction of A$_V$\,$\sim$\,\magap{27} and a 
dereddened absolute M$_{\rm H}$ of $-\magnit{1}{6}$, consistent with a 
10\Msolar{} ZAMS star with a spectral type of B1 (Schmidt-Kaler 1982). However, 
integrating under near-, mid-, and far-infrared fluxes, TSH02 derived a 
luminosity of $\sim$\,200\Lsolar, equivalent to only a 4\Msolar{} ZAMS star and 
closer to late~B in spectral type. TSH02 also found no evidence for the 
coincident Pa$\alpha$ emission from ionised gas that would be expected from 
even a late~B star, and thus, as they note, it is possible that YSO1 is perhaps 
a compact cluster or, in our opinion more likely, a young protostar. The 
2.7\,mm continuum emission from YSO1 reported by S02 might, for example, be 
free-free emission from an associated ultracompact \HII{} region rather than 
a disk, perhaps supporting the young, high-mass protostar hypothesis.

\begin{table}
\caption{Photometry and derived properties for sources embedded in the 
trunks and EGGs assuming 1\,Myr ages for all sources}
\begin{center}
\begin{tabular}{lcccccc}
\hline
Source$^{\rm a}$           & 
\filter{J$_s$}$^{\rm b}$   & 
\filter{H}$^{\rm b}$       & 
\filter{K$_s$}$^{\rm b}$   & 
Mass$^{\rm c}$             & 
A$_V$$^{\rm c}$            & 
Excess$^{\rm d}$           \\
 & & & & [\Msolar] & [mag] & \\ 
\hline
YSO1   & \magnit{17}{8}&\magnit{14}{7}& \magnit{12}{9}& 10     &  27   &  Yes \\
YSO2   & \magnit{15}{2}&\magnit{13}{8}& \magnit{12}{8}& 2--6   &  15   &  Yes \\
E14    & \magnit{20}{0}&\magnit{18}{8}& \magnit{17}{7}& 0.04   &   7   &  Yes \\
E15    & \magnit{22}{5}&\magnit{20}{7}& \magnit{19}{8}& 0.03   &  12   &  No  \\
E17    & \magnit{20}{7}&\magnit{19}{9}& \magnit{18}{6}& 0.02   &   3   &  Yes \\
E18    & \magnit{19}{3}&\magnit{19}{5}& \magnit{18}{8}& (0.04) &  (9)  &  Yes \\
E19    & \magnit{19}{1}&\magnit{19}{0}& \magnit{18}{2}& (0.05) &  (5)  &  Yes \\
E24    & \magnit{20}{7}&\magnit{19}{7}& \magnit{18}{4}& 0.02   &   5   &  Yes \\
E25    & \magnit{16}{9}&\magnit{15}{3}& \magnit{13}{8}& 0.5    &   9   &  Yes \\
E31    & \magnit{17}{6}&\magnit{15}{9}& \magnit{14}{6}& 0.35   &  10   &  Yes \\
E35    & \magnit{20}{0}&\magnit{17}{1}& \magnit{15}{6}& 0.95   &  22   &  No  \\
E42    & \magnit{14}{4}&\magnit{13}{4}& \magnit{12}{9}& 1.0    &   4   &  No  \\
E57 & $>$\magnit{23}{0}&\magnit{21}{1}& \magnit{19}{3}& (0.07) & (25)  &  No  \\
\hline
\end{tabular}
\end{center}
$^{\rm a}$YSO1 and YSO2 are the sources embedded in the tips of C1 and C2 as
   described in the text; E$nn$ are EGGs (see Fig.\,8 of H96) \\
$^{\rm b}$Photometry in 0.45$^{\prime\prime}$ diameter aperture, with aperture 
   correction \\
$^{\rm c}$Mass and extinction were calculated from \filter{J$_s$} and 
   \filter{H}, apart from those in parentheses which used \filter{H} 
   and \filter{K$_s$} \\
$^{\rm d}$Excess long-wavelength emission 
   in the \colour{J$_s$}{H} vs.\ \colour{H}{K$_s$} diagram
\end{table}

The tip of C2 also contains a bright point source (here YSO2; T1 of S02, 
M16ES-2 of TSH02) illuminating a compact reflection nebulosity. S02 claim that 
this nebulosity is bipolar, but instead we see a fainter second source roughly 
1~arcsec south, and which we find to be consistent with a lightly reddened 
field star (see also Fig.\,10b of H96). Again assuming photospheric emission 
for YSO2, we deduce an extinction of $\sim$\magap{15} and a mass of 
$\sim$\,2--5\Msolar. The non-monotonic mass-luminosity relation for young, 
higher-mass sources yields this ambiguity for YSO2, but we note that the 
luminosity of $\sim$\,20\Lsolar{} derived by TSH02 corresponds to a 
2--3\Msolar{} star at 1\,Myr (Palla \& Stahler 1999).

By contrast, C3 has no embedded sources in its tip, although a pair of 
relatively bright sources sit in the `bowl' just above it. TSH02 suggest that 
these may be young stars recently revealed as the tip of C3 was evaporated 
away, but from our wider-field images, it appears equally plausible they are 
simply field stars: the chances of such a pair being projected there is small,
but not vanishingly so. Finally, to the south-east of the HST trunks and 
outside the image shown here (see McCaughrean \& Andersen 2001), there is 
another similar structure that has not yet been overrun by the ionisation 
front. The VLT images show an extended red reflection nebula in the tip of 
this column, which Andersen \etal{} (2002) have suggested to be the driving 
source of the Herbig-Haro flow HH\,216, part of which is seen as the blue 
nebulous arc near the base of Fig.\,1.

\subsubsection{Star formation in the EGGs}
The key issue here, previously addressed briefly by McCaughrean (1997),
is to assess statistically whether or not the EGGs, as a population of 
small dense cores, are indeed forming young stars. To test this hypothesis, we 
first registered the HST optical and VLT near-infrared frames using 60 stars 
common to both: the fit has an RMS error of 0.02 arcsec, accurate enough 
given the median EGG diameter of 0.5 arcsec. Using a list of coordinates and 
dimensions supplied by Jeff Hester, we examined each EGG in the HST H$\alpha$ 
data, the individual VLT infrared mosaics, and the infrared colour composite. 
This process is subjective to an extent, as was the original definition of 
the EGGs: some are well defined and partly or completely detached from the 
surface of the trunks, but most are mere protrusions above the local surface.
In these cases, we made a rough judgement of the size of the underlying 
EGG based on its apparent radius of curvature, and then searched within
those boundaries.

In the 73 EGGs, we find 11 that show definite evidence for an associated 
near-infrared point source (Table~1). Two of these (E31, E42) sit right at the
tips of `mini-trunks' near the base of C1 and are optically visible, as
noted by H96, but the remaining 9 (E14, E15, E17, E18, E19, E24, E25, 
E35, E57) are only visible at near-infrared wavelengths. Using the
method and assumptions given above, we find that four of the sources (E25, 
E31, E35, E42) appear to have stellar masses in the range 0.3--1\Msolar, while 
the remainder (E14, E15, E17, E18, E19, E24, E57) are substellar, with masses 
in the range 0.02--0.07\Msolar. Remarkably, most of these substellar sources 
(E14--E19) are associated with rather small EGGs located at the tip of C1, 
near the much more massive YSO1. The apparent lack of low-mass stars in this 
region may simply be a selection effect: somewhat higher-mass sources evolve 
more rapidly and may have already escaped their EGGs. In this case, we would 
not have picked them out against the field stars according to the present 
EGG-based search criteria.

Another 5 EGGs show tentative evidence for embedded sources (E16, E20, E45, 
E46, E60), while a further 6 EGGs show near-infrared sources nearby, but on 
various grounds we consider these to be unrelated (E03, E23, E44, E50, E64). 
The remaining 51 EGGs show no evidence for associated sources in our data, 
including E01, which H96 suggested was coincident with a bright near-infrared 
source in the data of Hillenbrand \etal{} (1993): the VLT data clearly show 
it to be a background field star lying off the EGG\@. A more detailed 
discussion of each EGG will be given in our subsequent paper.

Given the huge density of background field stars in the region, some sources 
apparently coincident with EGGs may simply be chance projections. We can 
approach this problem in two ways. First, plain field stars should have normal 
reddened near-infrared colours, while young low-mass sources should show 
infrared excess emission typical of a circumstellar disk. The sources 
associated with E14, E17, E18, E19, E24, E25, and E31 do indeed show 
infrared excess confirming them as young, but the absence of excess emission 
from the other candidates should not necessarily be taken as evidence against 
them, as only half of all YSOs in Taurus-Auriga with excess 10\micron{} 
emission also show it in their near-infrared colours (Kenyon \& Hartmann 1995). 
Longer wavelength high spatial resolution imaging would provide crucial 
additional confirmation, although the required sensitivity may mean waiting 
until the NGST is available.

Second, we have taken advantage of the wide field covered by our data to 
examine the field population away from the trunks and count the number of 
stars per magnitude bin per unit area. For each star seen against an EGG, we 
can then calculate the chances of there being a field star projected within the 
confines of the EGG at that measured magnitude, after accounting for the 
additional extinction introduced by the EGG\@. We find this correction to be 
negligibly small in almost all cases, as will be discussed in our subsequent 
paper.

\section{Conclusions}
Our VLT images of the M\,16 elephant trunks show clear evidence for recent 
star formation. There are relatively massive, embedded YSOs at the tips of 
two of the HST columns, as also discussed by S02 and TSH02, and in one 
further column to the south-east. Furthermore, the VLT data have been able
to demonstrate that while the majority of the Eagle's EGGs are sterile, at 
least some are fertile. That is, some 15\% of them appear to be real `eggs', 
hatching new stars or brown dwarfs. This fraction is no larger than that
found by McCaughrean (1997), despite the significantly improved data. The
main reason is that several previously-claimed coincidences disappeared
as the spatial resolution improved, only to be replaced by new discoveries
at higher sensitivity.

This 15\% must still be considered a lower limit however, as some of the EGGs 
remain opaque even at near-infrared wavelengths. A case in point is E23, seen 
against the darkest part of C1 (Fig.\,1). There is no associated near-infrared 
point source, but both infrared and optical images show a linear jet-like 
feature extending to the north, ending in a nebular knot, with a counter 
`bowshock' an equal distance to the south. This circumstantially implies the 
presence of a very deeply embedded protostar within E23, driving an outflow. 
Also, there may be other young stars in and around the columns unrelated to 
EGGs. Infrared excess emission can be used to identify sources of this kind: 
S02 found several additional candidate YSOs in this manner, and we shall return 
to this question in our subsequent paper.

Finally, for all our focus on the trunks and EGGs, we should keep in mind the 
wider context, in particular the adjacent cluster NGC\,6611, which has formed 
thousands of stars within the past few million years (Hillenbrand \etal{} 
1993). The ongoing destruction of the columns and the relatively limited star 
formation taking place within them may ultimately prove to be a sideshow in 
the grander scheme of things, albeit a beautiful one.

\begin{acknowledgements}
We thank our friends at the VLT who took these excellent data in service mode; 
Jeff Hester for supplying the original HST images and EGG coordinates; Marc 
Gagn\'e for interesting discussions on M\,16; Francesco Palla and Isabelle 
Baraffe for providing their PMS tracks; and Sandra Jorge and Sofia Fernandes 
for their independent assessment of the EGGs. This work is supported by the EC 
RTN ``The Formation and Evolution of Young Stellar Clusters'' 
(HPRN-CT-2000-00155) and DLR grants 50-OR-0004/9912.
\end{acknowledgements}

\end{document}